\documentstyle[11pt,axodraw,fleqn]{article}
\newcounter{abc}
\def\bitem{\begin{itemize}}
\def\eitem{\end{itemize}}
\def\benum{\begin{enumerate}}
\def\eenum{\end{enumerate}}
\def\bdesc{\begin{description}}
\def\edesc{\end{description}}
\def\be {\begin{eqnarray}}
\def\ee {\end{eqnarray}}
\def\beq {\begin{equation}}
\def\eeq {\end{equation}}
\def\fr{\frac}
\def\ie {{\it i.e.}}
\def\etc{{\it etc.}}

\def\eg {{\it e.g.}}

\def\pr {Phys. Rev.}
\def\rmp {Rev. Mod. Phys.}
\def\jmp {J. Math. Phys.}

\def\ctpb {Comm. Theor. Phys. ( Beijing )}

\def\bi {\begin{itemize}}
\def\ei {\end{itemize}}
\def\ben {\begin{enumerate}}
\def\een {\end{enumerate}}

\begin{document}

\hfill{\bf AS-ITP-95-24}

\hfill{\it August, 1995}

\begin{center}
\vskip 0.4in

{\Large\bf {Intrinsic Regularization Method in QCD }\footnote{\sf Work
supported in part by the National Natural Science Foundation of China.}}\\

\vskip 0.4in

{\large\sf Yu Cai \footnote{\sf Email: caiyu@itp.ac.cn} ~and~ Han-Ying Guo
\footnote{\sf Email: hyguo@itp.ac.cn}}\\

\vskip 0.2in

{\sf Institute of Theoretical Physics,  Academia Sinica,}\\
{\sf P.O.Box 2735, Beijing 100080, P.R.China.}\\

\vskip 0.35in

{\large\sf Dao-Neng Gao}\\
\vskip 0.2in

{\sf Fundamental Physics Center, University of Science \& Technology of
China,}\\
{\sf Hefei Anhui, 230026, P.R.China.}\\

\vskip 0.6in


\begin{minipage}{5in}

\centerline{\Large\sf Abstract}

\vskip 0.5cm
\noindent
{\it There exist certain intrinsic relations between the ultraviolet divergent
graphs and the convergent ones at the same loop order in  renormalizable
quantum field theories. Whereupon we may establish a new method, the intrinsic
regularization method, to regularize those divergent graphs.  In this paper, we
apply this method to QCD at the one loop order. It turns out to be
satisfactory: The gauge invariance is preserved manifestly and the results are
the same as those derived by means of other regularization methods. }

\bigskip

\end{minipage}
\end{center}

\par\vfill

\newpage

\setcounter{footnote}{0}
\section{Introduction}
\indent
Over the decades, as is well known, a wide variety of regularization schemes
have been developed in quantum field theory \cite{1}. However, as every schemes
heve their own distinct advantages and disadvantages, this topic is still one
of the important and fundamental issues under investigation. One of the most
challenging problem is perhaps how to preserve all properties of the original
action manifestly and consistently.

A few years ago, a new regularization method named intrinsic vertex
regularization was first proposed for the $\phi^4$ theory by Wang and Guo
\cite{2}. The key point of the method is, in fact,
based upon the following simple observation: For a given  ultraviolet
divergent function at certain loop order in a renormalizable QFT, there always
exists a set of convergent functions at the same loop order such that their
Feynman graphs share the same loop skeleton  and the main difference is that
the convergent ones have additional vertices of certain kind and the original
one is the case without these vertices. This is, in fact, a certain intrinsic
relation between the original  ultraviolet  divergent graph and the convergent
ones in the QFT. It is this relation that indicates it is possible to introduce
the regularized function for the divergent function with the help of those
convergent ones so that the potentially divergent integral of the graph can be
rendered  finite while for the limiting case of the number of the additional
vertices $q\to 0$ the divergence again becomes manifest in pole(s) of $q$.

To be concrete, let us consider a 1PI graph with $I$ internal lines at one loop
order in the $\phi^4$ theory. Its superficial degree of divergences in the
momentum  space is
$$
\delta=4-2I.
$$
When $I=1$ or $2$, the graph is divergent. Obviously, there exists such kind of
graphs that they have additional $q$ four-$\phi$-vertices in the internal
lines. Then the number of internal lines in these graphs is $I+q$  so that the
divergent degree of the new 1PI graphs become
$$
\delta'=4-2(I+q).
$$
If $q$ is large enough, the new ones are convergent and the original divergent
one is the case of $q=0$.  Thus, a certain intrinsic relation has been reached
between the  original divergent 1PI graph and the new convergent ones at the
same loop order.

However, application of this method to QED runs into a difficulty. The problem
is that, unlike the the $\phi^4$ theory, the electron-photon vertex in QED
carries a $\gamma$-matrix and is a Lorentz vector. As a result, simply
inserting the vertex would increase the rank of the function as Lorentz tensors
and would make the problem quite complicated. In
order to overcome this difficulty, in \cite{3,4} the authors introduced an
alternative method. We follow the example of the $\phi^4$ theory to demonstrate
the key point of this method: One shifts the mass term of
the $\phi$ field from $m^2$ to $m^2+\mu^2$ and regards $\mu^2\phi^2$ as a new
vertex in addition to the vertex $\lambda\phi^4$. By inserting  the new vertex
into the internal $\phi$ lines in the graph of a given 1PI $n$-point divergent
function, a set of new convergent functions can be obtained provided the number
of inserted vertices, $q$, is large enough. Then one can introduce a new
convergent function, the regularized function, and the potential infinity in
the original 1PI $n$-point function may be recovered as the $q\to 0$ limiting
case of that function. Obviously, the mass shifting method can be easily
generalized to QED by simply shifting the electron mass from $m$ to $m+\mu$ and
regarding the term $-\mu\bar{\psi}{\psi}$ as a new vertex. In fact, as has been
shown in \cite{4}, it turns out to be successful to QED. Nevertheless, it is
not really
intrinsic since the Feynman rules of the theory have to be modified. As a
result, it may not completely work for non-Abelian gauge theories, \eg, QCD,
because generally it is not clear whether the gauge symmetry can be preserved
manifestly for these theories, although such a proof for QED at one loop level
has been given \cite{6}.

Very recently, we presented an improved approach in \cite{7} to reexamine the
$\phi^4$ theory and QED, in which a new concept, {\it inserter}, was
introduced. An inserter is a vertex or a pair of vertices linked by an internal
line, in which the momenta of the external legs are all set to zero, and, if
there are any, all the Lorentz indices are contracted in pair by the spacetime
metric and all the internal gauge symmetry indices are contracted by the
Killing-Cartan metric in the corresponding representation, so that as a whole
an inserter always carries the vacuum quantum numbers, i.e. zero momentum,
scalar in the spacetime symmetry, and singlet in internal and gauge symmetries.
It is not hard to see that in any given QFT as long as a suitable kind of
inserters are constructed with the help of the Feynman rules of the theory,
some intrinsic relations between the divergent functions and convergent ones at
the same loop order will be found by simply regarding the convergent ones as
the ones given by suitably

 inserting $q$-inserters in all internal lines in the given divergent ones.
The crucial point of this approach, therefore, is very simple but fundamental,
that is, the entire procedure is intrinsic in the QFT. There is nothing
changed, the action, the Feynman rules, the spacetime dimensions etc. are all
the same as that in the given QFT. This is a very important property which
should shed light on the challenging problem mentioned at the beginning of the
paper. Consequently, in applying to other cases all symmetries and topological
properties there should be preserved  in principle.

In what follows, we concentrate on how to apply the inserter approach to QCD at
one loop order. We present the main steps and the results of the inserter
regularization procedure for it. We find that, as is expected, the gauge
invariance is preserved manifestly, and all results are the same as those
derived by means of other regularization methods.

\section{Intrinsic Regularization in QCD}
\indent

 The QCD Lagrangian, including ghost fields and gauge fixing terms, can be
written as
\be
{\cal L}_{QCD} & = & -\frac{1}{4}
(\partial_{\mu} A^a_{\nu}-\partial_{\nu}A^a_{\mu}+g_cf^{abc}A^b_{\mu}A^c_{\nu})
(\partial^{\mu} A^{a\nu} -\partial^{\nu} A^{a\mu}+g_cf^{ade}A^{d\mu}A^{e\nu})
 \nonumber \\
  &  & -\frac{1}{2\xi}(\partial_{\mu} A^{a\mu})^2 -\bar{\eta}^a
\partial_{\mu}(\partial^{\mu} \delta^{ac}-g_cf^{abc}A^{b\mu})\eta^c
+\bar{\psi}[i\gamma_{\mu}(\partial^{\mu}
- ig_cA^{a\mu} \frac{\lambda^a}{2})-m]\psi~,
\label{L} \ee
and the Feynman rules are well known.

The main steps of the inserter approach for QCD may be stated more concretely
as follows. First, we should construct the inserters in QCD. This work, with
regards to simplicity and consistency with other theories, \eg, the
electroweak theory, may actually be done within a more general framework,
namely, within the framework of the standard model in which QCD is contained.
The explicit expressions of all inserters in the standard model have been
preestablished in \cite{7}. Here we merely list those relevant to QCD:

\bitem
\item
The gluon-inserter:
\be
I^{\{g\}ab}_{~~\mu\nu}(p)=-6ig_c^2C_2({\bf 8}) g_{\mu\nu} \delta^{ab}.
\label{gluon-ins} \ee

\item
The ghost-inserter:
\be
I^{\{gh\}}_{~~~a_1a_2}(p) = -ig_c^2 C_2({\bf 8}) \delta_{a_1a_2}.
\label{ghost-ins} \ee

\item
The quark-inserter:
\be
I^{\{q\}}(p)=-i\lambda_q.
\label{quark-ins} \ee
\eitem

In eqs.(\ref{gluon-ins}) and (\ref{ghost-ins}), $C_2({\bf 8})$ is the second
Casimir operator valued in the adjoint representation of $SU_c(3)$ algebra.
In eq.(\ref{quark-ins}), $\lambda_q$ takes value $\frac{g}{2}\frac{m_q}{M_W}$
in the standard model, but here its value is irrelevant for our purpose. It
should be mentioned that here the quark inserters are constructed by borrowing
the fermion-Higgs-vertex of Yukawa type from the standard model, this is in
analogy with as occurs in QED. The issue has been discussed in detail in
\cite{7}.

For a given  divergent 1PI amplitude
$\Gamma^{(n_f,n_g)}(p_1,\cdots, p_{n_f};k_1,\cdots, k_{n_g})$
at the one loop order with $n_f$ external fermion lines and $n_g$ external
photon lines, we consider a set of 1PI amplitudes
$\Gamma^{(n_f,n_g)} (p_1,\cdots, p_{n_f};k_1,\cdots, k_{n_g}; q)$
which correspond to the graphs with, if the loop contained in the graph purely
consists of fermion lines, all possible $2q$ insertions of the fermion inserter
in the internal fermion lines, or in other cases, all possible $q$ insertions
of
the corresponding inserter in the internal boson (ghost) lines in the original
graph. The divergent degree therefore becomes:
$$
\delta=4-I_f-2I_g-2q.
$$
If $q$ is large enough,
$\Gamma^{(n_f,n_g)}(p_1,\cdots,p_{n_f};k_1,\cdots, k_{n_g}; q)$
are convergent  and the original divergent function is the case of $q=0$. Thus
we reach a relation between the given divergent 1PI function and a set of
convergent 1PI functions at the one loop order. In fact, the function of
inserting the inserter(s) into internal lines is simply to raise the power of
the propagator of the lines and to decrease the degree of divergence of given
graph.

In order to regularize the given divergent function with the help of this
relation, we need to deal with those convergent functions on an
 equal footing and pay attention to their differences due to the insertions.
To this end, we introduce a new function:
\be
\begin{array}{ll}
\Gamma^{(n_f,n_g)}(p_1,\cdots, p_{n_f};k_1,\cdots, k_{n_g}; ~q;~\mu)
{}~~~~~~~\\[4mm]
{}~~~~~~~= (-i\mu)^{2q} (-i\lambda)^{-2q} \fr 1 {N_q} \sum
\Gamma^{(n_f,n_g)}(p_1,\cdots, p_{n_f};k_1,\cdots, k_{n_g}; ~q)
\end{array}
\ee
where $\mu$ is an arbitrary reference mass parameter,
the summation is taken over the entire set of such ${N_q}$ inserted functions,
and the factor $(-i\lambda)^{-2q}$ introduced here, in which $\lambda$ stands
for $\lambda_q$ for fermion loop and for $g_c$ for other cases, is to cancel
the ones coming from the inserters. It is clear that this function is the
arithmetical average of  those convergent functions and has  the same dimension
in mass, the same order in coupling constant  with the original divergent 1PI
function. Then we evaluate it and analytically continue $q$ from the integer to
the complex number. Finally, the  original 1PI function is recovered as its
$q\to 0$ limiting case:
\be
\Gamma^{(n_f,n_g)}(p_1,\cdots, p_{n_f};k_1,\cdots, k_{n_g})
=\lim_{q\to 0}
\Gamma^{(n_f,n_g)}(p_1,\cdots, p_{n_f};k_1,\cdots, k_{n_g};q;\mu),
\ee
and the original infinity appears as pole in $q$. Similarly, this procedure
should work for the cases at the higher loop orders in principle.

The divergent 1PI graphs at the one loop order in QCD are as follows: the gluon
self-energy $\Pi_{\mu \nu}^{ab}(k)$, the quark self-energy $\Sigma(p)$,  the
ghost self-energy $\tilde{\Pi}^{ab}(p)$, renormalized by $Z_3$, $Z^F_3$ and
$\tilde{Z}_3$, the three-gluon vertex $\Gamma^{abc}_{\mu\nu\lambda}
(k_1,~k_2)$, the four-gluon vertex
$\Gamma^{abcd}_{\mu\nu\lambda\tau}(k_1,~k_2,~k_3)$, the quark-gluon vertex
$\Gamma^{a}_{\mu}(p^{\prime},p)$, and the ghost-gluon vertex
$\tilde{\Gamma}^{abc}_{\mu}(p^{\prime},p)$, with the renormalization constant
$Z_1$, $Z_4$, $Z^F_1$, $\tilde{Z}_1$. In addition, there is a mass shift for
the quark, which we shall ignore. All corresponding graphs are listed in
figures 1-7. As numerious diagrams are concerned, evaluating them one by one in
detail would be much lengthy and unnecessary. In the next section, we will
evaluate in detail the gluon self-energy $\Pi_{\mu \nu}^{ab}(k)$ as a typical
example to show the main step of the approach, paying special attentions to the
gauge inva

riance of the function. Then in the subsequent section, we will directly give
all the results corresponding to other involved diagrams to verify the
Slavnov-Taylor identities at one loop order.

\section{Regularization and Evaluation of the Gluon Self-Energy $\Pi_{\mu
\nu}^{ab}(k)$}
\indent

Before the detailed evaluations are presented, we should first refer to a
special problem which arises in any massless theories, \ie, the genuine
infrared divergence in these theories. In the regularization schemes, this
problem usually appears as the lack of consistent definitions of the
regularized Feynman integrals for the ones which are both ultraviolet and
infrared divergent. For instance, in the dimensional regularization scheme,
let's consider the massive integral
\be
\int\fr{d^{2\omega}l}{(2\pi)^{2\omega}}\frac{1}{(l^2+m^2)^n}
=\fr{i\Gamma (n-\omega)}{(4\pi)^{\omega}\Gamma (n)(m^2)^{n-\omega}}
\equiv I(m,~\omega,~n)~, ~~~(~m^2\neq 0~)
\label{massive} \ee
which converges for $\omega$ complex; the parameter $n$ is arbitrary but fixed.
We note first of all that the limit
$\displaystyle\lim_{m^2\to0}I(m,~\omega,~n)$ may or may not exist, depending on
the relative magnitudes of $n$ and $\omega$. But even if it did exist, another
problem could arise as we approach four-space (provided the original integral
is infrared divergent to begin with), because in general
$$
\displaystyle\lim_{\omega\to2} [ \displaystyle\lim_{m^2\to0} I(m,~\omega,~n)]
\neq \displaystyle\lim_{m^2\to0} [\displaystyle\lim_{\omega\to2}
I(m,~\omega,~n)],
$$
so that the massless integral $\displaystyle\lim_{\omega \to
2}{\int\fr{d^{2\omega}l}{(2\pi)^{2\omega}}\frac{1}{(l^2)^n}}$
can not be derived unambiguously from the massive integral (\ref{massive}).
Furthermore, the trick of inserting a finite mass into the integral and then
allowing it to approach to zero at the end of the calculation is, in general,
not a satisfactory prescription yet, because it spoils the gauge symmetry in
the original theory, provided such a symmetry existed in the first place. To
avoid this difficulty, 't Hooft and Veltman naively comjectured that
\be
\displaystyle\lim_{\omega \to 2}{\int\fr{d^{2\omega}l}{(2\pi)^{2\omega}}
\frac{1}{(l^2)^n}}=0~,~~~for~ \omega,~n~complex.
\label{massless} \ee
It has been shown that no inconsistencies occur, \eg, in the Slavnov-Taylor
identities \cite{10}, due to the acceptance of the above conjecture.

In our application of the present approach to QCD, as we will see, the same
problem occurs, \eg, in calculating the tadpole diagram Fig.1d of the gluon
self-energy. To solve this problem, we employ a conjecture analogous to 't
Hooft and Veltman's:
\be
\displaystyle\lim_{q\to 0}{\int\fr{d^4l}{(2\pi)^4}
\frac{1}{[(k-l)^2]^{Aq}(l^2)^{Bq+n}}}=0,~~for~q,~n,~complex,~A\geq 0,~B\geq
0,~A+B=1~.
\label{conj} \ee
Likewise, we will see that no inconsistencies occur due to the acceptance of
the eq.(\ref{conj}).

Now we turn to evaluate in detail the gluon self-energy $\Pi_{\mu\nu}^{ab}(k)$
shown in Fig.1. The diagrams contributing to $\Pi_{\mu\nu}^{ab}(k)$ are four in
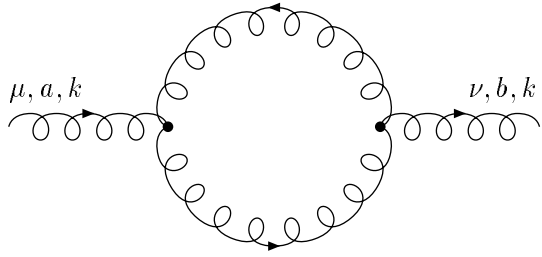
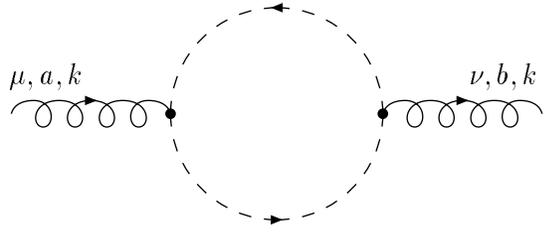
number, namely, the gluon loop diagram, the ghost loop diagram, the quark loop
diagrams, and the gluon tadpole.  The integral expressions of the regularized
diagrams in the momentum space are given in the appendix ( For simplicity, we
take the Feynman gauge $\xi =1$. ).

First, we consider the gluon loop contribution. From (\ref{glse-a}), a little
bit of algebra yields
\be
\Pi_{(A)\mu\nu}^{~~~ab}(k;q;\mu)~=~-\fr{1}{2}g_c^{2}[-6C_2({\bf 8})\mu^{2}]^q
     \delta^{ab}I_1 ~,  \nonumber
\ee
with
\be
I_1~=~\fr{1}{N_q} \displaystyle{\sum_{i=0}^{q} \int\fr{d^4p}{(2\pi)^4}}
\fr{-2p^2g_{\mu\nu}-(5k^2-2p\cdot k)g_{\mu\nu}
-10p_{\mu}p_{\nu}+2k_{\mu}k_{\nu}+5p_{\mu}k_{\nu}+5k_{\mu}p_{\nu}}{
(p^2)^{i+1}[(p-k)^2]^{q-i+1}}~.  \nonumber
\ee
In the present case, $N_q=q+1$.
Note that because of eq.(\ref{conj}), the contribution of the first term in the
numerator of the above equation actually vanishes, so it can be neglected.
Using the Feynman parameterization, we get
\be
I_1 & = & \fr{1}{q+1} \displaystyle{\sum_{i=0}^{q}} \fr{\Gamma (q+2)}{\Gamma
(i+1)\Gamma (q-i+1)}
\displaystyle{\int_0^1dx} x^{q-i}(1-x)^i   \nonumber  \\
 & &  \times
\displaystyle{\int\fr{d^4p}{(2\pi)^4}} \fr{(2x-5)k^2g_{\mu\nu}
-10p_{\mu}p_{\nu} -(10x^2-10x-2)k_{\mu}k_{\nu}}{[p^2+x(1-x)k^2]^{q+2}}~,
\label{I1}\ee
where we have made a momentun shift: $p \rightarrow p-kx$, and the linear terms
in $p$ in the numerator have been dropped since they do not contribute to the
integral. Now the integration over $p$ can be performed by using the following
formulas:
\renewcommand{\theequation}{\arabic{equation}\alph{abc}}
\setcounter{abc}{0}
\addtocounter{abc}{1}
\be
\int \fr{d^4 p}{(2\pi)^4} \fr{(p^2)^{\beta}}{(p^2+M^2)^A} & = &
\fr{i}{(4\pi)^2}
\fr{\Gamma(2+\beta)\Gamma(A-2-\beta)}{\Gamma(A)}(M^2)^{2+\beta-A},
 \label{int-a} \ee
\addtocounter{equation}{-1}
\addtocounter{abc}{1}
\vspace{-6mm}
\be
\int \fr{d^4 p}{(2\pi)^4} \fr{(p^2)^{\beta}p_{\mu}p_{\nu}}{(p^2+M^2)^A}
 & = &
\fr{i}{(4\pi)^2}\fr{1}{4}g_{\mu\nu}
\fr{\Gamma(3+\beta)\Gamma(A-3-\beta)}{\Gamma(A)}(M^2)^{3+\beta-A},
 \label{int-b} \ee
\addtocounter{equation}{-1}
\addtocounter{abc}{1}
\vspace{-6mm}
\be
\int \fr{d^4 p}{(2\pi)^4} \fr{(p^2)^{\beta}p_{\mu}p_{\nu}p_{\rho}p_{\sigma}}
{(p^2+M^2)^A}
& = &
\fr{i}{(4\pi)^2}\fr{1}{24}(g_{\mu\nu}g_{\rho\sigma}+g_{\mu\rho}g_{\nu\sigma}
+g_{\mu\sigma}g_{\nu\rho})  \nonumber \\
 & & \times
\fr{\Gamma(4+\beta)\Gamma(A-4-\beta)}{\Gamma(A)}(M^2)^{4+\beta-A}.
\label{int-c} \ee
\addtocounter{abc}{1}
\renewcommand{\theequation}{\arabic{equation}}
{}From (\ref{I1}), (\ref{int-a}), and (\ref{int-b}) we get:
$$
\begin{array}{ll}
I_1~ = & \fr{i}{(4\pi)^2} \displaystyle{\sum_{i=0}^{q}} \fr{\Gamma
(q-1)}{(q+1)\Gamma (i+1)\Gamma (q-i+1)} \displaystyle{\int_0^1dx}
x^{q-i}(1-x)^i   \nonumber  \\
  &  \times
\displaystyle{ \fr{(q-1)[(2x-5)k^2g_{\mu\nu}-(10x^2-10x-2)k_{\mu}k_{\nu}]
-5x(1-x)k^2g_{\mu\nu}}{[x(1-x)k^2]^{q}} }  \nonumber \\
 ~~~~~ = &
\fr{i}{(4\pi)^2} \fr{\Gamma (q-1)}{\Gamma (q+2)} \displaystyle{\int_0^1dx}
\fr{[q(2x-5)+(5x^2-7x+5)]k^2g_{\mu\nu}-(q-1)(10x^2-10x-2)k_{\mu}k_{\nu}}
{[x(1-x)k^2]^{q}}~,  \nonumber
\end{array}
$$
where in the last step the summation over $i$ has been performed with the help
of the binomial theorem. This expression also makes sense when we make an
analytical continuation of $q$ from integer to complex number. When $q\to 0$,
using the expansion
$$
\fr{\Gamma(q-1)}{\Gamma(q+2)}=-\fr{1}{q}-q+O(q^2),~~~
[x(1-x)k^2]^q=1+q\ln [x(1-x)k^2]+O(q^2)~,
$$
and making a rescaling of the parameter $\mu^2 \rightarrow -6C_2({\bf
8})\mu^2$,  we get:
\be
\Pi_{(A)\mu\nu}^{~~~ab}(k;q;\mu) & = & \fr{ig_c^2C_2({\bf 8})}{(4\pi)^2}
\delta^{ab} [ (\fr{19}{12}k^2g_{\mu\nu}-\fr{11}{6}k_{\mu}k_{\nu})\fr{1}{q}
\nonumber \\
 & &
-(\fr{19}{12}k^2g_{\mu\nu}-\fr{11}{6}k_{\mu}k_{\nu})\ln\Bigl(\fr{k^2}{\mu^2}\Bigr)+(\fr{47}{36}k^2g_{\mu\nu}-\fr{14}{9}k_{\mu}k_{\nu}) ]~.
\label{no13} \ee
Clearly $\Pi_{(A)\mu\nu}^{~~~ab}(k;q;\mu)$ does not conserve current, this is
due to our choice to use a covariant gauge (rather than, for example, an axial
gauge). For the sake of this choice, we have to introduce spurious gluon
polarization states. These spurious states must be removed by taking the ghost
loop contribution into account. We could have computed with an axial or ``ghost
free'' gauge, but it is usually much easier to use the simple Feynman gauge and
add in the ghost contribution.

To compute ghost loop contribution (\ref{glse-b}), we use the same Feynman
parameterization arriving at
\be
\Pi_{(B)\mu\nu}^{~~~ab}(k;q;\mu) ~=~ -g_c^2 C_2({\bf 8})[C_2({\bf 8})\mu^2]^q
\delta^{ab}I_2 ~,  \nonumber
\ee
with
\be
I_2 ~=~ \displaystyle{{\int_0^1dx} \int\fr{d^4p}{(2\pi)^4}}
\fr{p_{\mu}p_{\nu}-x(1-x)k_{\mu}k_{\nu}}{[p^2+x(1-x)k^2]^{q+2}}~, \nonumber
\ee
where momentun shift: $p \rightarrow p-kx$ has been made, and the linear terms
in $p$ in the numerator have been dropped. After performing the integration
over $p$ and taking the limit $q\to 0$ subsequently, we obtain
\be
\Pi_{(B)\mu\nu}^{~~~ab}(k;q;\mu)& = & \fr{ig_c^2C_2({\bf 8})}{(4\pi)^2}
\delta^{ab} [ (\fr{1}{12}k^2g_{\mu\nu}+\fr{1}{6}k_{\mu}k_{\nu})\fr{1}{q}
\nonumber \\
 & &
-(\fr{1}{12}k^2g_{\mu\nu}-\fr{1}{6}k_{\mu}k_{\nu})\ln\Bigl(\fr{k^2}{\mu^2}\Bigr)+\fr{5}{36}k^2g_{\mu\nu}+\fr{1}{9}k_{\mu}k_{\nu}) ]~.
\label{no14} \ee
Again, we find that $\Pi_{(B)\mu\nu}^{~~~ab}(k;q;\mu)$ also does not conserve
current. However, the non-conserving term in (\ref{no14}) exactly cancels that
in (\ref{no13}), and makes
$\Pi_{(A)\mu\nu}^{~~~ab}(k;q;\mu)+\Pi_{(B)\mu\nu}^{~~~ab}(k;q;\mu)$ gauge
invariant, namely,
$$
k^{\mu}[\Pi_{(A)\mu\nu}^{~~~ab}(k;q;\mu)+\Pi_{(B)\mu\nu}^{~~~ab}(k;q;\mu)]
=0~.
$$

Next, the quark loop contribution $\Pi_{(C)\mu\nu}^{~~~ab}(k;q;\mu)$ need not
be recalculated since it is just $\Pi_{\mu\nu}^{QED}(k;q;\mu)$, which can be
found in \cite{6,7}, multiplied by a color factor $N_fTr({\bf T}^a{\bf T}^b)$:
\be
\Pi_{(C)\mu\nu}^{~~~ab}(k;q;\mu) & = & N_fTr({\bf T}^a{\bf T}^b)
\Pi_{\mu\nu}^{QED}(k;q;\mu)=\frac{4ig_c^2N_fC_2({\bf 3})}{(4\pi)^2} \delta^{ab}
(k_{\mu}k_{\nu}-k^2g_{\mu\nu}) \nonumber \\
 & & \times
[\frac{1}{3q}+\frac{2}{3}-2\int_{0}^{1} dx x(1-x)\ln\frac{k^2 x(1-x)-m^2}
 {\mu^2}+\cdots]~.
\label{no15} \ee
Obviously, $\Pi_{(C)\mu\nu}^{~~~ab}(k;q;\mu)$ itself is gauge invariant.

Finally, as in dimensional regularization scheme, from eq.(\ref{conj}) we know
that the contribution of the gluon tadpole diagram vanishes:
\be
\Pi_{(D)\mu\nu}^{~~~ab}(k;q;\mu)=0~.
\label{no16} \ee

{}From (\ref{no13}), (\ref{no14}), (\ref{no15}), and (\ref{no16}) we obtain the
final expression of the regularized gluon self-energy
$\Pi_{\mu\nu}^{ab}(k;q;\mu)$:
\be
\Pi_{\mu\nu}^{ab}(k;q;\mu) & = & \Pi_{(A)\mu\nu}^{~~~ab}(k;q;\mu)
+\Pi_{(B)\mu\nu}^{~~~ab}(k;q;\mu)+\Pi_{(C)\mu\nu}^{~~~ab}(k;q;\mu) \nonumber \\
 & = & \frac{ig_c^2}{(4\pi)^2}\delta^{ab}(k^2g_{\mu\nu}-k_{\mu}k_{\nu})
\{[\fr{5}{3}C_2({\bf 8})-\fr{4}{3}N_fC_2({\bf 3})]\fr{1}{q}+\cdots\}~,
\ee
which is the same as that derived by means of other regularization methods and
satisfies the gauge invariance condition $k^{\mu}\Pi_{\mu\nu}^{ab}(k;q;\mu)=0$.

\section{Renormalization Constants and Slavnov-Taylor Identities at
one Loop Order}
\indent

So far we have calculated the gluon self-energy $\Pi_{\mu\nu}^{ab}(k)$ in
detail by means of the intrinsic regularization method at one loop level. As we
have expected, the result turns out to be gauge invariant. However, this is not
enough for us to say that the method preserves the gauge invariance of QCD. As
a complement, we should further show that the Slavnov-Taylor identities between
the renormalization constants hold, namely,
\be
\fr{Z_1}{Z_3} ~=~ \fr{Z^F_1}{Z^F_3} ~=~ \fr{\tilde{Z}_1}{\tilde{Z}_3} ~=~
\fr{Z_4}{Z_1}~.
\label{stid} \ee
It is these identities that guarantee that the renormalized theory possess the
same gauge theory structure as the original one. Moreover, they are essential
for proving the renormalizability and unitarity of the theory.

To verify these identities, all divergent 1PI graphs, not only
$\Pi_{\mu\nu}^{ab}(k)$, but also others, must be taken into account. After
lengthy and tedious calculations, we get

\renewcommand{\theequation}{\arabic{equation}\alph{abc}}
\setcounter{abc}{0}
\addtocounter{abc}{1}
\vspace{-6mm}
\be
Z_1=1+\fr{g_c^2}{(4\pi)^2}[\fr{2}{3}C_2({\bf 8})-\fr{4}{3}N_fC_2({\bf
3})]\fr{1}{q}
\label{const-a} \ee
\addtocounter{equation}{-1}
\addtocounter{abc}{1}
\vspace{-6mm}
\be
Z_3=1+\fr{g_c^2}{(4\pi)^2}[\fr{5}{3}C_2({\bf 8})-\fr{4}{3}N_fC_2({\bf
3})]\fr{1}{q}
\label{const-b} \ee
\addtocounter{equation}{-1}
\addtocounter{abc}{1}
\vspace{-6mm}
\be
Z_4=1+\fr{g_c^2}{(4\pi)^2}[-\fr{1}{3}C_2({\bf 8})-\fr{4}{3}N_fC_2({\bf
3})]\fr{1}{q}
\label{const-c} \ee
\addtocounter{equation}{-1}
\addtocounter{abc}{1}
\vspace{-6mm}
\be
Z^F_1=1-\fr{g_c^2}{(4\pi)^2}[C_2({\bf 8})+\fr{8}{3}C_2({\bf 3})]\fr{1}{q}
\label{const-d} \ee
\addtocounter{equation}{-1}
\addtocounter{abc}{1}
\vspace{-6mm}
\be
Z^F_3=1-\fr{g_c^2}{(4\pi)^2}\fr{8}{3}C_2({\bf 3})\fr{1}{q}
\label{const-e} \ee
\addtocounter{equation}{-1}
\addtocounter{abc}{1}
\vspace{-6mm}
\be
\tilde{Z}_1=1-\fr{g_c^2}{(4\pi)^2}\fr{C_2({\bf 8})}{2}\fr{1}{q}
\label{const-f} \ee
\addtocounter{equation}{-1}
\addtocounter{abc}{1}
\vspace{-6mm}
\be
\tilde{Z}_3=1+\fr{g_c^2}{(4\pi)^2}\fr{C_2({\bf 8})}{2}\fr{1}{q}
\label{const-g} \ee
\addtocounter{abc}{1}
\renewcommand{\theequation}{\arabic{equation}}

It is easy to see that the Slavnov-Taylor identities (\ref{stid}) indeed holds.
For a close observation, we note that although these expressions are calculated
in a specific gauge therefore their gauge dependence are not explicit, they are
in fact all gauge dependent ( For instance, in the axial gauge, the effective
Lagrantian has no ghosts and has the same structure as in QED, resulting in the
identity $Z^F_1=Z^F_3$, which is patently untrue in the Feynman gauge.). We
also remark that the fermion contribution to the vector boson quartic
self-interaction must diverge if the Slavnov-Taylor identities (\ref{stid}) are
to hold because we explicitly see that the ratio $Z^F_1/Z^F_3Z_3^{\fr{1}{2}}$,
by which the bare coupling constant is related to the renormalized  one,
contains a fermion contribution. On the contrary, the corresponding box diagram
of QED is finite. The reason is that in QED the box diagram's divergence
vanishes only upon symmetrization of the external photon lines denoted only by
their vector in

dices, while in QCD the symmetrization of the external lines can be performed
in two ways: by symmetrizing on both vector and group indices, which as in QED
gives no divergence, {\it or} by antisymmetrizing on both vector and group
indices. It is this new contribution which deverges. The same reason can be
applied to the fermion contribution to the triple gauge vertex.

\section{Concluding Remarks}
\indent

We have shown the main steps and results for  the regularization of the
divergent 1PI functions at the one loop order in QCD by means of the inserter
proposal for the intrinsic regularization method. It turns out to be
satisfactory: The gauge invariance is preserved manifestly and the results are
the same as those derived by means of other regularization methods. It is
natural to expect that this proposal should  be available to the cases at
higher loop orders in principle.

The renormalization of the QCD under consideration in this scheme should be the
same as in  usual approaches. Namely, we may subtract the divergent part of the
$n$-point functions at each loop order by adding the relevant counterterms to
the action. The renormalized $n$-point functions are then evaluated from the
renormalized action. In the limiting case, we get the finite results for all
correlation functions.

It should be mentioned that the method presented here is somewhat analogous to
the analytic regularization method developed by Speer \cite{11,12}. However,
the two methods are in fact different: As is well known, the analytic
regularization method violates unitarity, this is due to the fact that it
actually continuing the power of propagators in an arbitrary way. While in our
method this not the case. Here we trie to find out a procedure of
regularization from some physical principles. That is, giving a ultraviolet
divergent process, one can always find a set of convergent function obtainable
from existing Feymann rules and for the limiting case it turns to be the
original ultra-divergent one. Or from another point of view, given a
ultra-divergent function, one can always extract a set of convergent functions
from a certain physical process, which is gauge invariant.
After taking the limitation of the convergent functions, the divergent
function is naturally regularized. There is nothing changed, the action, the
Feynman rules, the spacetime dimensions etc. are all the same as that in the
given QFT. From this viewpoint, one should have no doubt of gauge invariance
and the unitarity of the method since the sum of the convergent functions comes
from a certain physical process.

Application of our approach to gauge theories containing spontaneous symmetry
breaking such as the standard model should be straightforward. Also, it will be
much helpful to apply our approach to some other cases, such as anomalies, SUSY
theories \etc, since in these cases the symmetries and topological properties
are sensitive to the spacetime dimensions and the number of fermionic degrees
of freedom \etc, thus we are unable to tackle them consistently by means of the
hitherto well-known regularization methods such as dimensional regularization
method. It is reasonable to expect that the approach presented here should be
able to get rid of those problems. We will investigate these issues in detail
elsewhere.

\bigskip
\bigskip

{\raggedright{\Large\bf Appendix: Integral Expressions of the Divergent 1PI
Graphs at}}\\
{\raggedright{\Large\bf \* \* \* \* \* \* \* \* \* \* \* \* \* \* \* One Loop
Level in QCD}}\\
\indent

There are a number of divergent iPI graphs at one loop level in QCD, which
contribute to the gluon self-energy $\Pi_{\mu \nu}^{ab}(k)$, the quark
self-energy $\Sigma(p)$, the ghost self-energy $\tilde{\Pi}^{ab}(p)$, the
three-gluon vertex $\Gamma^{abc}_{\mu\nu\lambda} (k_1,~k_2)$, the four-gluon
vertex $\Gamma^{abcd}_{\mu\nu\lambda\tau}(k_1,~k_2,~k_3)$, the quark-gluon
vertex $\Gamma^{a}_{\mu}(p^{\prime},p)$, and the ghost-gluon vertex
$\tilde{\Gamma}^{abc}_{\mu}(p^{\prime},p)$ respectively. In what follows we
present the integral expressions of the regularized diagrams in the momentum
space (in Feynman gauge $\xi =1$).

\noindent
1. \hspace{0.3cm}The integral expressions of the regularized diagrams
contributing to $\Pi_{\mu \nu}^{ab}(k)$:
\renewcommand{\theequation}{\arabic{equation}\alph{abc}}
\setcounter{abc}{0}
\addtocounter{abc}{1}
\be
\Pi_{(A)\mu\nu}^{~~~ab}(k;q;\mu)& = & \fr{1}{2}g_c^{2-2q}
\mu^{2q}[-6ig_c^2C_2({\bf 8})]^q Tr({\bf F}^a{\bf F}^b)
\fr{1}{N_q}\sum_{i=0}^{q} \int\fr{d^4p}{(2\pi)^4}  \nonumber \\
  &  &  \times
G_{\mu\rho_1\sigma_1}(k,-p,p-k) G_{\nu\rho_2\sigma_2}(-k,p,k-p) \nonumber \\
  &  &  \times
g^{\rho_1\rho_2} g^{\sigma_1\sigma_2} \Bigl(\fr{-i}{p^2}\Bigr)^{i+1}
\Bigl(\fr{-i}{(p-k)^2}\Bigr)^{q-i+1}~,
 \label{glse-a} \ee
\addtocounter{equation}{-1}
\addtocounter{abc}{1}
\be
\Pi_{(B)\mu\nu}^{~~~ab}(k;q;\mu)& = & g_c^{2-2q} \mu^{2q}[-ig_c^2C_2({\bf
8})]^q Tr({\bf F}^a{\bf F}^b) \fr{1}{N_q}\sum_{i=0}^{q} \int\fr{d^4p}{(2\pi)^4}
 \nonumber \\
 & & \times p_{\mu}(p-k)_{\nu}\Bigl(\fr{-i}{p^2}\Bigr)^{i+1}
\Bigl(\fr{-i}{(p-k)^2}\Bigr)^{q-i+1}~,
 \label{glse-b} \ee
\addtocounter{equation}{-1}
\addtocounter{abc}{1}
\be
\Pi_{(C)\mu\nu}^{~~~ab}(k;q;\mu)& = & g_c^{2}\mu^{2q} (-i\lambda_q)^{2q} N_f
Tr({\bf T}^a{\bf T}^b)\fr{1}{N_q}\sum_{i=0}^{2q} \int\fr{d^4p}{(2\pi)^4}
\nonumber \\
 & & \times Tr[\gamma_{\mu}\Bigl(\fr{1}{\not{p}-\not{k}-m}\Bigr)^{i+1}
\gamma_{\nu}\Bigl(\fr{1}{ \not {p} -m}\Bigr)^{2q-i+1}]~,
 \label{glse-c} \ee
\addtocounter{equation}{-1}
\addtocounter{abc}{1}
\be
\Pi_{(D)\mu\nu}^{~~~ab}(k;q;\mu)& = &
-\fr{1}{2}ig_c^{2-2q}\mu^{2q}[-6ig_c^2C_2({\bf 8})]^q [f^{abe}f^{cde}
(g_{\mu\sigma}g_{\nu\rho}-g_{\mu\rho}g_{\nu\sigma}) \nonumber \\
 & & +f^{ace}f^{dbe}(g_{\mu\rho}g_{\nu\sigma}-g_{\mu\nu}g_{\rho\sigma})+
f^{ade}f^{bce}(g_{\mu\nu}g_{\rho\sigma}-g_{\mu\sigma}g_{\nu\rho})] \nonumber \\
 & & \times g^{\sigma\rho}\delta^{cd} \int\fr{d^4p}{(2\pi)^4}
\Bigl(\fr{-i}{p^2}\Bigr)^{q+1}~,
 \label{glse-d} \ee
\addtocounter{abc}{1}
\renewcommand{\theequation}{\arabic{equation}}
where $N_f$ is the number of flavors, ${\bf T}^a$ are generators of $SU_c(3)$
in fundamental representation, $f^{abc}$ denote the structure constants of
$SU_c(3)$, $({\bf F}^a)^{bc}=-if^{abc}$ are generators of $SU_c(3)$ in adjoint
representation, and
$$
G_{\mu\nu\lambda}(p_1,p_2,p_3)~=~(p_1-p_2)_{\lambda}g_{\mu\nu}+(p_2-p_3)_{\mu}g_{\nu\lambda}+(p_3-p_1)_{\nu}g_{\lambda\mu}
$$
comes from the three-gluon vertex.


\bigskip
\noindent
2. \hspace{0.3cm}The integral expressions of the regularized diagrams
contributing to $\Gamma^{abc}_{\mu\nu\lambda} (k_1,~k_2)$:
\renewcommand{\theequation}{\arabic{equation}\alph{abc}}
\setcounter{abc}{0}
\addtocounter{abc}{1}
\begin{equation}
\begin{array}{l}
\displaystyle{\Gamma_{(A)\mu\nu\lambda}^{~~~abc}(k_1,~k_2;q;\mu) ~ = ~
ig_c^{3-2q} \mu^{2q}[-6ig_c^2C_2({\bf 8})]^q Tr({\bf F}^a{\bf F}^b{\bf F}^c)
\fr{1}{N_q}\sum_{i=0}^{q} \sum_{j=0}^{q-i} \int\fr{d^4p}{(2\pi)^4} }  \nonumber
\\
{}~~~~~~~~~~~~~ \times
\displaystyle{g^{\rho_1\sigma_2} g^{\rho_2\sigma_3} g^{\rho_3\sigma_1}
G_{\nu\rho_2\sigma_2}(k_2,-k_2-p,p)
G_{\lambda\rho_3\sigma_3}(-k_1-k_2,k_1-p,k_2+p)  } \nonumber \\
{}~~~~~~~~~~~~~  \times
\displaystyle{G_{\mu\rho_1\sigma_1}(k_1,-p,p-k_1)
\Bigl(\fr{-i}{p^2}\Bigr)^{i+1} \Bigl(\fr{-i}{(p+k_2)^2}\Bigr)^{j+1}
\Bigl(\fr{-i}{(p-k_1)^2}\Bigr)^{q-i-j+1}  }~,
\end{array}
\label{3g-a} \end{equation}
\addtocounter{equation}{-1}
\addtocounter{abc}{1}
\begin{equation}
\begin{array}{l}
\displaystyle{
\Gamma_{(B)\mu\nu\lambda}^{~~~abc}(k_1,~k_2;q;\mu) ~ = ~ -\fr{1}{2}ig_c^{3-2q}
\mu^{2q}[-6ig_c^2C_2({\bf 8})]^q \fr{1}{N_q}\sum_{i=0}^{q}
\int\fr{d^4p}{(2\pi)^4}f^{as_1t_1}  } \nonumber \\
{}~~~~~~~~~~~~~  \displaystyle{ G_{\mu\rho_1\sigma_1}(k_1,-p,p-k_1)
[f^{bcr}f^{t_2s_2r}(g_{\nu\sigma_2}g_{\lambda\rho_2}-g_{\nu\rho_2}g_{\lambda\sigma_2}) }  \nonumber \\
{}~~~~~~~~~~~~~
\displaystyle{+f^{bt_2r}f^{s_2cr}(g_{\nu\rho_2}g_{\lambda\sigma_2}-g_{\nu\lambda}g_{\rho_2\sigma_2})+f^{bs_2r}f^{ct_2r}(g_{\nu\lambda}g_{\rho_2\sigma_2}-g_{\nu\sigma_2}g_{\lambda\rho_2})] } \nonumber \\
{}~~~~~~~~~~~~~  \times
\displaystyle{\delta^{s_1s_2} \delta^{t_1t_2} g^{\rho_1\rho_2}
g^{\sigma_1\sigma_2} \Bigl(\fr{-i}{p^2}\Bigr)^{i+1}
\Bigl(\fr{-i}{(p-k_1)^2}\Bigr)^{q-i+1} }  \nonumber \\
{}~~~~~~~~~~~~~
+\{(\mu,a,k_1) \leftrightarrow (\nu,b,k_2)\}+\{(\mu,a,k_1)\leftrightarrow
(\lambda,c,k_3)\}~,
\end{array}
\label{3g-b}  \end{equation}
\addtocounter{equation}{-1}
\addtocounter{abc}{1}
\begin{equation}
\begin{array}{l}
\displaystyle{
\Gamma_{(C)\mu\nu\lambda}^{~~~abc}(k_1,~k_2;q;\mu) ~ = ~ -ig_c^{3-2q}
\mu^{2q}[-ig_c^2C_2({\bf 8})]^q Tr({\bf F}^a{\bf F}^b{\bf F}^c)
\fr{1}{N_q}\sum_{i=0}^{q} \sum_{j=0}^{q-i} \int\fr{d^4p}{(2\pi)^4}  }
\nonumber \\
{}~~~~~~~~~~~~~  \times
\displaystyle{p_{\mu}(p+k_2)_{\nu}(p-k_1)_{\lambda}
\Bigl(\fr{i}{p^2}\Bigr)^{i+1}\Bigl(\fr{i}{(p+k_2)^2}\Bigr)^{j+1}
\Bigl(\fr{i}{(p-k_1)^2}\Bigr)^{q-i-j+1} } \nonumber \\
{}~~~~~~~~~~~~~
+\{(\nu,b,k_2) \leftrightarrow (\lambda,c,k_3)\}~,
\end{array}
\label{3g-c}  \end{equation}
\addtocounter{equation}{-1}
\addtocounter{abc}{1}
\begin{equation}
\begin{array}{l}
\displaystyle{
\Gamma_{(D)\mu\nu\lambda}^{~~~abc}(k_1,~k_2;q;\mu)  ~ = ~ ig_c^{3}\mu^{2q}
(-i\lambda_q)^{2q} Tr({\bf T}^a{\bf T}^b{\bf T}^c) \fr{1}{N_q}  \sum_{i=0}^{2q}
\sum_{j=0}^{2q-i} \int\fr{d^4p}{(2\pi)^4} }~~~~~~~~~~ \nonumber\\
{}~~~~~~~~~~~~~  \times
\displaystyle{Tr[ \gamma_{\mu}\Bigl(\fr{i}{\not{p}-\not{k}_1-m}\Bigr)^{i+1}
\gamma_{\lambda}\Bigl(\fr{i}{\not{p}+\not{k}_2-m}\Bigr)^{j+1}
\gamma_{\nu}\Bigl(\fr{i}{\not{p}-m}\Bigr)^{2q-i-j+1}] } \nonumber \\
{}~~~~~~~~~~~~~
+\{(\nu,b,k_2) \leftrightarrow (\lambda,c,k_3)\}~,
\end{array}
\label{3g-d}  \end{equation}
\addtocounter{abc}{1}
\renewcommand{\theequation}{\arabic{equation}}

\bigskip
\noindent
3.  \hspace{0.3cm}The integral expressions of the regularized diagrams
contributing to $\Gamma^{abcd}_{\mu\nu\lambda\tau}(k_1,~k_2,~k_3)$:
\renewcommand{\theequation}{\arabic{equation}\alph{abc}}
\setcounter{abc}{0}
\addtocounter{abc}{1}
\begin{equation}
\begin{array}{l}
\displaystyle{
\Gamma_{(A)\mu\nu\lambda\tau}^{~~~~abcd}(k_1,~k_2,~k_3;q;\mu) ~ = ~ g_c^{4-2q}
\mu^{2q}[-6ig_c^2C_2({\bf 8})]^q Tr({\bf F}^a{\bf F}^b{\bf F}^c{\bf F}^d)
\fr{1}{N_q}\sum_{i=0}^{q} \sum_{j=0}^{q-i} \sum_{l=0}^{q-i-j}  } \nonumber \\
{}~~~~~~~~~~~~~    \times
\displaystyle{
\int\fr{d^4p}{(2\pi)^4} G_{\mu\rho_1\sigma_1}(k_1,-p,p-k_1)
G_{\nu\rho_2\sigma_2}(k_2,-k_2-p,p)   } \nonumber \\
{}~~~~~~~~~~~~~   \times
\displaystyle{ G_{\tau\rho_4\sigma_4}(-k_1-k_2-k_3,k_1-p,k_2+k_3+p)   }
\nonumber  \\
{}~~~~~~~~~~~~~   \times
\displaystyle{  G_{\lambda\rho_3\sigma_3}(k_3,-k_2-k_3-p,k_2+p)
 g^{\rho_1\sigma_2} g^{\rho_2\sigma_3} g^{\rho_3\sigma_4} g^{\rho_4\sigma_1}  }
\nonumber \\
{}~~~~~~~~~~~~~   \times
\displaystyle{
\Bigl(\fr{-i}{p^2}\Bigr)^{i+1} \Bigl(\fr{-i}{(p+k_2)^2}\Bigr)^{j+1}
\Bigl(\fr{-i}{(p+k_2+k_3)^2}\Bigr)^{l+1}\Bigl(\fr{-i}{(p-k_1)^2}\Bigr)^{q-i-j-l+1} } \nonumber \\
{}~~~~~~~~~~~~~
+\{(\nu,b,k_2) \rightarrow (\lambda,c,k_3),~(\lambda,c,k_3) \rightarrow
(\tau,d,k_4),~(\tau,d,k_4) \rightarrow (\nu,b,k_2)\} \nonumber \\
{}~~~~~~~~~~~~~
+\{(\nu,b,k_2) \rightarrow (\tau,d,k_4)),~(\tau,d,k_4) \rightarrow
(\lambda,c,k_3),~(\lambda,c,k_3) \rightarrow (\nu,b,k_2)\}~,
\end{array}
\label{4g-a}  \end{equation}
\addtocounter{equation}{-1}
\addtocounter{abc}{1}
\begin{equation}
\begin{array}{l}
\displaystyle{
\Gamma_{(B)\mu\nu\lambda\tau}^{~~~~abcd}(k_1,~k_2,~k_3;q;\mu) ~ = ~ -\fr{1}{2}
g_c^{4-2q} \mu^{2q}[-6ig_c^2C_2({\bf 8})]^q  \fr{1}{N_q}\sum_{i=0}^{q}
\int\fr{d^4p}{(2\pi)^4} }  \nonumber \\
{}~~~~~~~~~~~~~  \times
\displaystyle{
[f^{adr_1}f^{t_1s_1r_1}(g_{\mu\sigma_1}g_{\tau\rho_1}-g_{\mu\rho_1}g_{\tau\sigma_1})+f^{at_1r_1}f^{s_1dr_1}(g_{\mu\rho_1}g_{\tau\sigma_1}-g_{\mu\tau}g_{\rho_1\sigma_1})   } \nonumber \\
{}~~~~~~~~~~~~~
\displaystyle{
+f^{as_1r_1}f^{dt_1r_1}(g_{\mu\tau}g_{\rho_1\sigma_1}-g_{\mu\sigma_1}g_{\tau\rho_1})][f^{bcr_2}f^{t_2s_2r_2}(g_{\nu\sigma_2} g_{\lambda\rho_2}- g_{\nu\rho_2}g_{\lambda\sigma_2}) } \nonumber \\
{}~~~~~~~~~~~~~
\displaystyle{
+f^{bt_2r_2}f^{s_2cr_2}(g_{\nu\rho_2}g_{\lambda\sigma_2}-g_{\nu\lambda}g_{\rho_2\sigma_2})+f^{bs_2r_2}f^{ct_2r_2}(g_{\nu\lambda}g_{\rho_2\sigma_2}-g_{\nu\sigma_2}g_{\lambda\rho_2})] } \nonumber\\
{}~~~~~~~~~~~~~  \times
\displaystyle{
\delta^{s_1s_2}\delta^{t_1t_2} g^{\rho_1\rho_2} g^{\sigma_1\sigma_2}
\Bigl(\fr{-i}{p^2}\Bigr)^{i+1}  \Bigl(\fr{-i}{(p+k_2+k_3)^2}\Bigr)^{q-i+1} }
\nonumber \\
{}~~~~~~~~~~~~~
+\{(\tau,d,k_4) \leftrightarrow (\lambda,c,k_3)\}+\{(\tau,d,k_4)
\leftrightarrow (\nu,b,k_2)\}~,
\end{array}
\label{4g-b}  \end{equation}
\addtocounter{equation}{-1}
\addtocounter{abc}{1}
\begin{equation}
\begin{array}{l}
\displaystyle{
\Gamma_{(C)\mu\nu\lambda\tau}^{~~~~abcd}(k_1,~k_2,~k_3;q;\mu) ~ = ~
-ig_c^{4-2q} \mu^{2q}[-6ig_c^2C_2({\bf 8})]^q  \fr{1}{N_q}\sum_{i=0}^{q}
\sum_{j=0}^{q-i} \int\fr{d^4p}{(2\pi)^4} } \nonumber \\
{}~~~~~~~~~~~~~  \times
\displaystyle{   \delta^{s_1t_2}\delta^{s_2t_3}\delta^{s_3t_1}
[f^{adr}f^{t_1s_1r}(g_{\mu\sigma_1}g_{\tau\rho_1}-g_{\mu\rho_1}g_{\tau\sigma_1})   } \nonumber \\
{}~~~~~~~~~~~~~
\displaystyle{
+f^{at_1r}f^{s_1dr}(g_{\mu\rho_1}g_{\tau\sigma_1}-g_{\mu\tau}g_{\rho_1\sigma_1})
+f^{as_1r}f^{dt_1r}(g_{\mu\tau}g_{\rho_1\sigma_1}-g_{\mu\sigma_1}g_{\tau\rho_1})
]   } \nonumber \\
{}~~~~~~~~~~~~~  \times
\displaystyle{  f^{bs_2t_2}f^{cs_3t_3}  G_{\nu\rho_2\sigma_2}(k_2,-k_2-p,p)
G_{\lambda\rho_3\sigma_3}(k_3,-k_2-k_3-p,k_2+p) } \nonumber \\
{}~~~~~~~~~~~~~  \times
\displaystyle{ g^{\rho_1\sigma_2} g^{\rho_2\sigma_3} g^{\rho_3\sigma_1}
\Bigl(\fr{-i}{p^2}\Bigr)^{i+1} \Bigl(\fr{-i}{(p+k_2)^2}\Bigr)^{j+1}
\Bigl(\fr{-i}{(p+k_2+k_3)^2}\Bigr)^{l+1}  } \nonumber \\
{}~~~~~~~~~~~~~
+\{(\tau,d,k_4) \leftrightarrow (\lambda,c,k_3)\}+\{(\tau,d,k_4)
\leftrightarrow (\nu,b,k_2)\} \nonumber \\
{}~~~~~~~~~~~~~
+\{(\mu,a,k_1) \leftrightarrow (\lambda,c,k_3)\}+\{(\mu,a,k_1) \leftrightarrow
(\nu,b,k_2)\} \nonumber \\
{}~~~~~~~~~~~~~
+\{(\mu,a,k_1) \leftrightarrow (\nu,b,k_2),~(\tau,d,k_4) \leftrightarrow
(\lambda,c,k_3)\}~,
\end{array}
\label{4g-c}  \end{equation}
\addtocounter{equation}{-1}
\addtocounter{abc}{1}
\begin{equation}
\begin{array}{l}
\displaystyle{
\Gamma_{(D)\mu\nu\lambda\tau}^{~~~~abcd}(k_1,~k_2,~k_3;q;\mu) ~ = ~ -g_c^{4-2q}
\mu^{2q}[-6ig_c^2C_2({\bf 8})]^q Tr({\bf F}^a{\bf F}^b{\bf F}^c{\bf F}^d)
\fr{1}{N_q}\sum_{i=0}^{q} \sum_{j=0}^{q-i} \sum_{l=0}^{q-i-j}  } \nonumber \\
{}~~~~~~~~~~~~  \times
\displaystyle{
\int\fr{d^4p}{(2\pi)^4} p_{\mu}(p+k_2)_{\nu}(p+k_2+k_3)_{\lambda}(p-k_1)_{\tau}
\Bigl(\fr{i}{p^2}\Bigr)^{i+1}  }  \nonumber\\
{}~~~~~~~~~~~~   \times
\displaystyle{
\Bigl(\fr{i}{(p+k_2)^2}\Bigr)^{j+1}
\Bigl(\fr{i}{(p+k_2+k_3)^2}\Bigr)^{l+1}
\Bigl(\fr{i}{(p-k_1)^2}\Bigr)^{q-i-j-l+1}  } \nonumber \\
{}~~~~~~~~~~~~
+\{(\nu,b,k_2) \rightarrow (\lambda,c,k_3),~(\lambda,c,k_3) \rightarrow
(\tau,d,k_4),~(\tau,d,k_4) \rightarrow (\nu,b,k_2)\} \nonumber \\
{}~~~~~~~~~~~~
+\{(\nu,b,k_2) \rightarrow (\tau,d,k_4)),~(\tau,d,k_4) \rightarrow
(\lambda,c,k_3),~(\lambda,c,k_3) \rightarrow (\nu,b,k_2)\} \nonumber \\
{}~~~~~~~~~~~~
+\{(\nu,b,k_2) \leftrightarrow (\lambda,c,k_3)\}+\{(\nu,b,k_2) \leftrightarrow
(\tau,d,k_4)\}+\{(\lambda,c,k_3) \leftrightarrow (\tau,d,k_4)\},
\end{array}
\label{4g-d}  \end{equation}
\addtocounter{equation}{-1}
\addtocounter{abc}{1}
\begin{equation}
\begin{array}{l}
\displaystyle{
\Gamma_{(E)\mu\nu\lambda\tau}^{~~~~abcd}(k_1,~k_2,~k_3;q;\mu) ~ = ~
-g_c^{4}\mu^{2q} (-i\lambda_q)^{2q} Tr({\bf T}^a{\bf T}^b{\bf T}^c{\bf T}^d)
\fr{1}{N_q}\sum_{i=0}^{2q} \sum_{j=0}^{2q-i} \sum_{l=0}^{2q-i-j}   }
\nonumber\\
{}~~~~~~~~~~~~  \times
\displaystyle{   \int\fr{d^4p}{(2\pi)^4}
Tr[\gamma_{\mu}\Bigl(\fr{i}{\not{p}-\not{k}_1-m}\Bigr)^{i+1}
\gamma_{\tau}\Bigl(\fr{i}{\not{p}+\not{k}_2+\not{k}_3-m}\Bigr)^{j+1} }
\nonumber\\
{}~~~~~~~~~~~~  \times
\displaystyle{
\gamma_{\lambda}\Bigl(\fr{i}{\not{p}+\not{k}_2-m}\Bigr)^{l+1}
\gamma_{\nu}\Bigl(\fr{i}{\not{p}-m}\Bigr)^{2q-i-j-l+1}]  }  \nonumber \\
{}~~~~~~~~~~~~
+\{(\nu,b,k_2) \rightarrow (\lambda,c,k_3),~(\lambda,c,k_3) \rightarrow
(\tau,d,k_4),~(\tau,d,k_4) \rightarrow (\nu,b,k_2)\} \nonumber \\
{}~~~~~~~~~~~~
+\{(\nu,b,k_2) \rightarrow (\tau,d,k_4)),~(\tau,d,k_4) \rightarrow
(\lambda,c,k_3),~(\lambda,c,k_3) \rightarrow (\nu,b,k_2)\} \nonumber \\
{}~~~~~~~~~~~~
+\{(\nu,b,k_2) \leftrightarrow (\lambda,c,k_3)\}+\{(\nu,b,k_2) \leftrightarrow
(\tau,d,k_4)\}+\{(\lambda,c,k_3) \leftrightarrow (\tau,d,k_4)\}~,
\end{array}
\label{4g-e}  \end{equation}
\addtocounter{abc}{1}
\renewcommand{\theequation}{\arabic{equation}}

\bigskip
\noindent
4.  \hspace{0.3cm}The integral expressions of the regularized diagram
contributing to $\tilde{\Pi}^{ab}(p)$:
\be
\tilde{\Pi}^{ab}(p;q;\mu) & = & -g_c^{2-2q} \mu^{2q}[-ig_c^2C_2({\bf 8})]^q
Tr({\bf F}^a{\bf F}^b) \int\fr{d^4k}{(2\pi)^4} p\cdot k
\Bigl(\fr{i}{k^2}\Bigr)^{q+1} \Bigl(\fr{-i}{(k-p)^2}\Bigr)~,
\label{ghse} \ee

\bigskip
\noindent
5.  \hspace{0.3cm}The integral expressions of the regularized diagrams
contributing to $\tilde{\Gamma}^{abc}_{\mu}(p^{\prime},p)$:
\renewcommand{\theequation}{\arabic{equation}\alph{abc}}
\setcounter{abc}{0}
\addtocounter{abc}{1}
\be
\tilde{\Gamma}^{~abc}_{(A)\mu}(p^{\prime},p;q;\mu) & = & ig_c^{3-2q}
\mu^{2q}[-ig_c^2C_2({\bf 8})]^q Tr({\bf F}^a{\bf F}^b{\bf F}^c)
\int\fr{d^4k}{(2\pi)^4} k^{\sigma}p^{\prime\rho}  \nonumber \\
 & & \times
G_{\mu\rho\sigma}(k_1-k_2,p-k_1,k_2-p)
\Bigl(\fr{i}{k^2}\Bigr)^{q+1}
\Bigl(\fr{-i}{(k-p^{\prime})^2}\Bigr)\Bigl(\fr{-i}{(k-p)^2}\Bigr)~,
\label{glgh-a} \ee
\addtocounter{equation}{-1}
\addtocounter{abc}{1}
\be
\tilde{\Gamma}^{~abc}_{(B)\mu}(p^{\prime},p;q;\mu) & = & ig_c^{3-2q}
\mu^{2q}[-6ig_c^2C_2({\bf 8})]^q Tr({\bf F}^a{\bf F}^b{\bf F}^c)
\displaystyle{\int\fr{d^4k}{(2\pi)^4}} (k+p^{\prime})_{\mu}(k+p)\cdot
p^{\prime} \nonumber \\
 & & \times
\displaystyle{\Bigl(\fr{-i}{k^2}\Bigr)^{q+1}
\Bigl(\fr{i}{(k+p^{\prime})^2}\Bigr)\Bigl(\fr{i}{(k+p)^2}\Bigr)}~,
\label{glgh-b} \ee
\addtocounter{abc}{1}
\renewcommand{\theequation}{\arabic{equation}}

\bigskip
\noindent
6.  \hspace{0.3cm}The integral expressions of the regularized diagram
contributing to  $\Sigma(p)$:
\be
\Sigma(p;q;\mu) & = & {\bf T}^a{\bf T}^a \Sigma^{QED}(p;q;\mu) \nonumber \\
 & = &
-g_c^{2-2q} \mu^{2q}[-6ig_c^2C_2({\bf 8})]^q {\bf T}^a{\bf T}^a
\int \fr{d^4 k}{(2\pi)^4}
(\gamma^{\mu}\fr{i}{ \not{k}-m}\gamma_{\mu})
\Bigl(\fr{-i}{(k-p)^2}\Bigr)^{q+1},
\label{qse} \ee

\bigskip
\noindent
7. \hspace{0.3cm}The integral expressions of the regularized diagrams
contributing to $\Gamma^{a}_{\mu}(p^{\prime},p)$:
\renewcommand{\theequation}{\arabic{equation}\alph{abc}}
\setcounter{abc}{0}
\addtocounter{abc}{1}
\be
\Gamma^{~~~a}_{(A)\mu}(p^{\prime},p;q;\mu) & = & -g_c^{3-2q} \mu^{2q}
[-6ig_c^2C_2({\bf 8})]^q {\bf T}^s{\bf T}^t \fr{1}{N_q}\sum_{i=0}^{q}
\int\fr{d^4k}{(2\pi)^4} \nonumber \\
 & & \times
f^{ast} G_{\mu\rho\sigma}(p^{\prime}-p,p-k,k-p^{\prime})  \nonumber \\
 & & \times
(\gamma^{\rho}\fr{i}{ \not{k}-m}\gamma^{\sigma})
\Bigl(\fr{-i}{(k-p)^2}\Bigr)^{i+1}\Bigl(\fr{-i}{(k-p^{\prime})^2}\Bigr)^{q-i+1},
\label{glq-a} \ee
\addtocounter{equation}{-1}
\addtocounter{abc}{1}
\be
\Gamma^{~~~a}_{(B)\mu}(p^{\prime},p;q;\mu) & = & {\bf T}^b{\bf T}^a{\bf T}^b
\Gamma^{QED}(p^{\prime},p;q;\mu) \nonumber \\
 & = &
 -ig_c^{3-2q} \mu^{2q}[-6ig_c^2C_2({\bf 8})]^q {\bf T}^b{\bf T}^a{\bf T}^b
\int \fr{d^4 k}{(2\pi)^4} \nonumber \\
 & & \times
(\gamma^{\rho}\fr{i}{\not{k}+\not{p^{\prime}}-m}
\gamma_{\mu}\fr{1}{\not{k}+\not{p}-m}
\gamma_{\rho}) \Bigl(\fr{-i}{k^2}\Bigr)^{q+1},
\label{glq-b} \ee
\addtocounter{abc}{1}
\renewcommand{\theequation}{\arabic{equation}}

\bigskip
\bigskip
{\raggedright{\Large \bf Acknowledgment}}
\vskip 0.5cm
\noindent
{The work is supported in part by the National Natural Science Foundation of
China. One of the author (YC) is also supported in part by Local Natural
Science Foundation of Xinjiang.}

\newpage

{\raggedright{\Large \bf Figure Captions}}
\bdesc

\vskip 1cm
\item{\bf Figure 1}
The one loop Feynman diagrams which are needed for calculation of the
renormalization constant $Z_3$.


\vskip 0.5cm
\item {\bf Figure 2}
The one loop Feynman diagrams which are needed for  calculation of the
renormalization constant $Z_1$.

\vskip 0.5cm
\item {\bf Figure 3}
The one loop Feynman diagrams which are needed for  calculation of the
renormalization constant $Z_4$.

\vskip 0.5cm
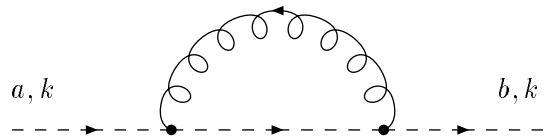
\item {\bf Figure 4}
The one loop Feynman diagram which is needed for  calculation of the
renormalization constant $\tilde{Z}_3$.

\vskip 0.5cm
\item {\bf Figure 5}
The one loop Feynman diagrams which are needed for  calculation of the
renormalization constant $\tilde{Z}_1$.

\vskip 0.5cm
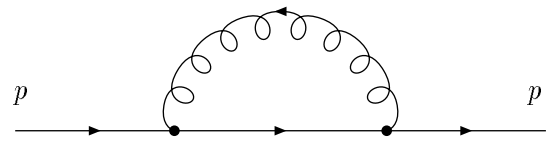
\item {\bf Figure 6}
The one loop Feynman diagram which is needed for  calculation of the
renormalization constant $Z^F_3$.

\vskip 0.5cm
\item {\bf Figure 7}
The one loop Feynman diagrams which are needed for  calculation of the
renormalization constant $Z^F_1$.

\edesc

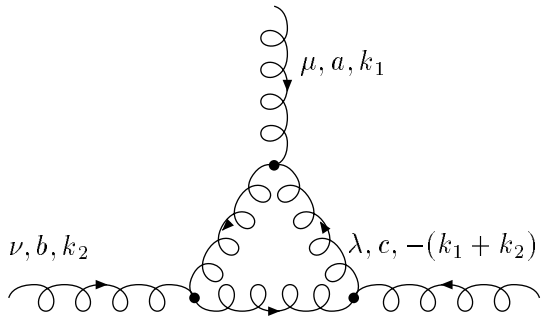
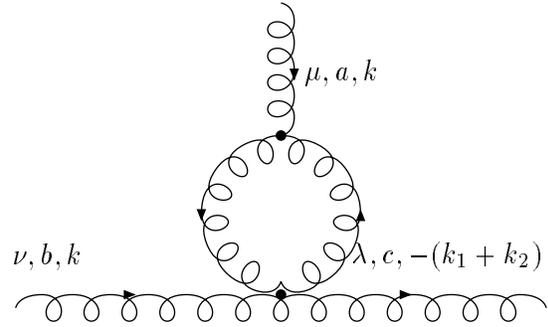
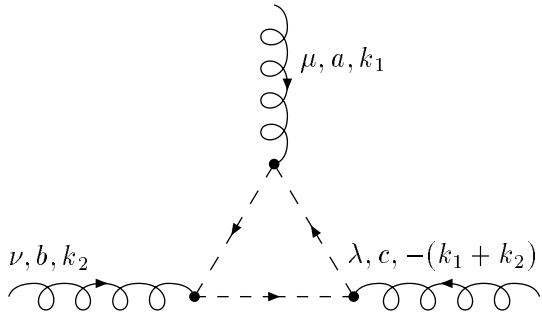
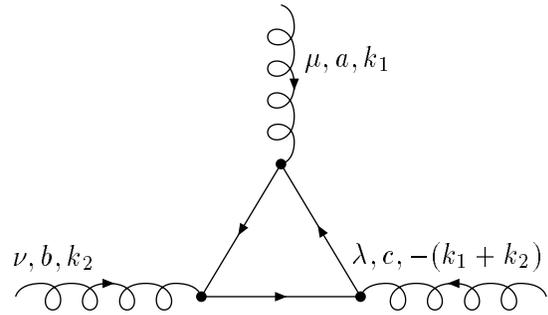
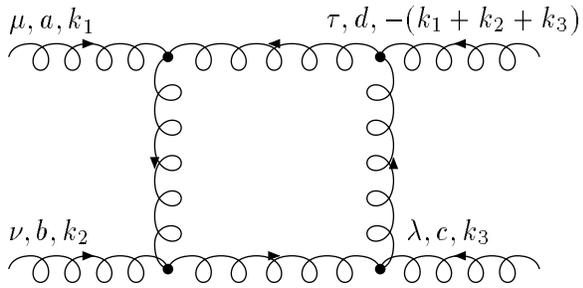
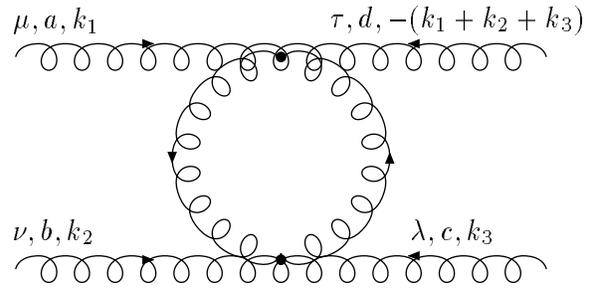
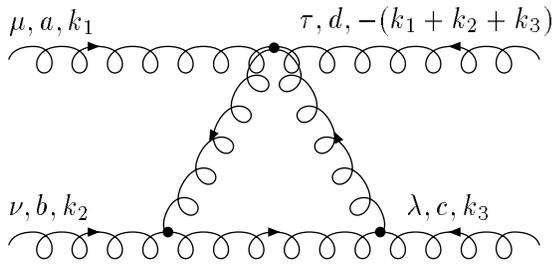
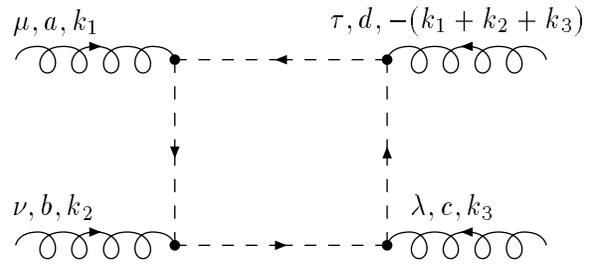
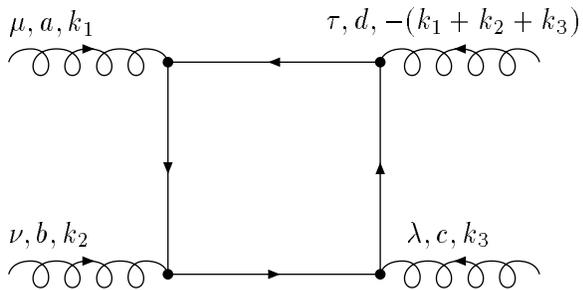
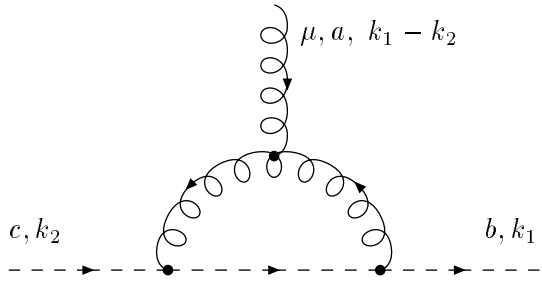
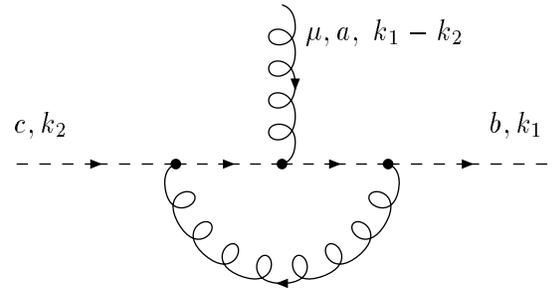
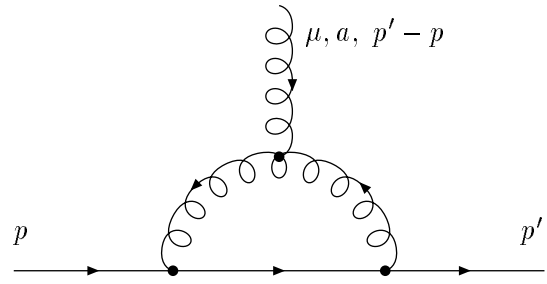
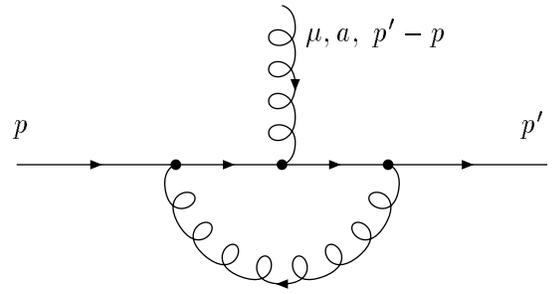
\end{document}